# Implementation of FAIR principles in the IPCC: The WGI AR6 Atlas repository

## Authors

Maialen Iturbide[1], Jesús Fernández[1], José M. Gutiérrez[1], Anna Pirani[2], David Huard[3], Alaa Al Khourdajie[4], Jorge Baño-Medina[1], Joaquin Bedia[5], Ana Casanueva[5], Ezequiel Cimadevilla[5], Antonio S. Cofiño[1], Matteo De Felice[6], Javier Diez-Sierra[1,5], Markel García-Díez[7], James Goldie[8], Dimitris A. Herrera[9], Sixto Herrera[5], Rodrigo Manzanas[5], Josipa Milovac[1], Aparna Radhakrishnan[10], Daniel San-Martín[7], Alessandro Spinuso[11], Kristen Thyng[12], Claire Trenham[13], Özge Yelekçi[14]

## Affiliations

1. Meteorology Group, Instituto de Física de Cantabria (IFCA), Universidad de Cantabria-CSIC. Santander, Spain.
2. International Centre for Theoretical Physics (ICTP), Italy
3. Ouranos, Montréal, Québec, Canada
4. Centre for Environmental Policy, Imperial College London, London, UK
5. Meteorology Group, Dept of Applied Mathematics and Computer Science, Universidad de Cantabria, Santander, Spain
6. European Commission, Joint Research Centre (JRC), Petten, The Netherlands
7. Predictia Intelligent Data Solutions S.L., Santander, Spain
8. 360info, Monash University, Melbourne, Australia
9. Instituto Geográfico Universitario, Universidad Autónoma de Santo Domingo, Santo Domingo 10103, Dominican Republic
10. Princeton University, Princeton, USA
11. Koninklijk Nederlands Meteorologisch Instituut (KNMI), De Bilt, The Netherlands
12. Axiom Data Science, USA
13. Commonwealth Scientific & Industrial Research Organisation (CSIRO) - Oceans & Atmosphere, Canberra, Australia
14. IPCC WGI TSU, Universite Paris Saclay, France

corresponding author(s): Jesús Fernández (Jesus.Fernandez@unican.es), José M. Gutiérrez (gutierjm@unican.es)

## Abstract

The Sixth Assessment Report (AR6) of the Intergovernmental Panel on Climate Change (IPCC) has adopted the FAIR Guiding Principles. The Atlas chapter of Working Group I (WGI) is presented as a test case. Here, we describe the application of these principles in the Atlas, the challenges faced during its implementation, and those that remain for the future. We present the open source repository resulting from this process, which collects the code (including annotated Jupyter notebooks), data provenance, and some aggregated datasets underpinning the key figures in the Atlas chapter and its interactive companion (the Interactive Atlas), open to scrutiny by the scientific community and the general public. We describe the informal pilot review conducted on this repository to gather recommendations that led to significant improvements. Finally, a working example illustrates the use of the repository to produce customized regional information, extending the Interactive Atlas products and running the code interactively in a web browser using Jupyter notebooks. Atlas repository: doi:10.5281/zenodo.5171760.

## Background & summary

The accessibility and reproducibility of scientific results is a major concern in all scientific disciplines[1]. This was identified as a key aspect to be addressed during the preparation of the Sixth Assessment Report (AR6) of the Intergovernmental Panel on Climate Change (IPCC) in order to ensure the transparency of the products underpinning the report and enhance their use. This was particularly relevant for the digital products developed as part of the Atlas chapter and the Interactive Atlas (https://interactive-atlas.ipcc.ch) of Working Group I (WGI). Best practices and standards for open science were defined and implemented in collaboration with the IPCC Task Group on Data Support for Climate Change Assessments (TG-Data, https://www.ipcc.ch/data). These are summarized in the TG-Data guidance document on FAIR (Findable, Accessible, Interoperable, Reusable) principles for IPCC AR6[2].

FAIR data principles[3,4] aim to facilitate open science by ensuring that the data and code used are findable and accessible and can be reused for reproducibility and for further developments using interoperable infrastructures. Guidance and training was provided to all authors, recommending that FAIR principles were adopted in the preparation of the WGI report. This is still ongoing work and has been successful to varying degrees, with data, metadata and scripts underpinning chapter and Technical Summary figures assembled. All the data underpinning the Summary for Policymakers (SPM) figures is publicly available[5]. These activities benefited from the adoption of documentation protocols, collaborative open-access frameworks (such as R or Python) and platforms (such as GitHub), which facilitated the collection of metadata (including data provenance) and code (including Jupyter notebooks), as well as some data, allowing the reproduction of key results and figures of the report. These activities are also being undertaken by the other working groups, namely WGII on climate adaptation, WGIII on climate mitigation and in the Synthesis Report (SR).

FAIR principles were particularly relevant for the Atlas chapter and the Interactive Atlas to facilitate reproducibility of complex figures based on multiple lines of evidence and reproducibility and reusability of the digital products and code, respectively. In particular, the Interactive Atlas is a novel online tool for flexible analyses of observed and projected climate change information for about 25 variables and indices underpinning the WGI contribution to AR6. The implementation of FAIR principles was particularly critical to develop a rigorous and transparent interactive tool as part of the WGI report, as well as an important contribution to a traceable implementation of the IPCC Error Protocol for report figures. Therefore, the Atlas is a comprehensive test case for the development and implementation of FAIR principles, exploring frameworks, protocols and best practices to be expanded to other chapters and the IPCC more broadly. For the Atlas chapter, these results have been collected and uploaded to an open source Atlas repository[6]. These have been made available for review and scrutiny of the scientific community and users more broadly, and this has improved the quality of the final Atlas products.

In this paper, we describe the FAIR guiding principles of the Atlas chapter repository, the informal review process in the context of the IPCC review process, and the challenges faced during their implementation and those remaining for the future. Overall, the informal review greatly improved the Atlas repository in all aspects and strengthened community trust in the available tools and products. We also provide an overview of the repository scope and contents, emphasizing the data and annotated notebooks readily available to reproduce Atlas figures. Finally, a new working example is provided to illustrate the use of the available resources to create and extend the Interactive Atlas products. In particular, we show how to create customized Global Warming Level (GWL) scaling plots using a well-documented Jupyter notebook and freely available online computational resources from MyBinder[7].

## Methods

The methodology followed during the implementation of FAIR principles in the Atlas is described here, as well as the Atlas GitHub repository that embodies these principles, with a special focus on reproducibility and reusability.

**The Atlas repository**

The Atlas repository contains the code (including Jupyter notebooks), data provenance and some aggregated datasets underpinning key figures of the Atlas chapter and its interactive companion (the Interactive Atlas). Figure 1 shows a schematic representation describing its structure and highlighting its contents. This repository is maintained on GitHub (https://github.com/IPCC-WG1/Atlas), which is the current *de facto* standard for code sharing with over 200 million repositories and over 65 million registered users. GitHub provides a collaborative environment for code development, with integrated version control and issue tracking among many other features. The repository is self-documented by Markdown-formatted files exploiting GitHub capabilities to browse folders and link to internal and external resources. Moreover, GitHub enables long-term archiving of repository snapshots, interoperating with services such as Zenodo (http://zenodo.org). The final release of the Atlas repository, containing data, metadata, code and documentation, is available not only through GitHub, as described next, but also as a Zenodo entry[6] with its own persistent Digital Object Identifier (DOI).

**Implementation of FAIR principles**

In the case of the WGI Atlas, FAIR principles are implemented through the Atlas repository described above. In particular:

**Findability** of the repository itself is promoted by its deployment on the widely-used GitHub platform. Additionally, Zenodo snapshots provide a unique, persistent identifier DOI. For all climate data inputs used in the Atlas, the repository lists the unique URLs (and DOIs, when available) for both data and metadata. This collection of unique identifiers is searchable, using either the automatic GitHub filter on comma-separated values (CSV) files or their version rendered as a searchable HTML table. Metadata are rich, both at the data file content description level (as enforced by the netCDF CF conventions (https://cfconventions.org) used for all data published on the Earth System Grid Federation (ESGF), as well as at the data provenance level (DOIs pointing to World Data Center for Climate -WDCC- entries describing model configuration and references).

**Accessibility** to the data and metadata in the Atlas repository is open. The full contents can be retrieved anonymously via HTTP, either from GitHub or Zenodo. All the content is licensed under a Creative Commons Attribution (CC-BY 4.0) license, except for some CORDEX datasets with non-commercial restrictions (CC-BY-NC 4.0 license). The original data and metadata sources listed are accessible through trusted repositories, in particular via authenticated HTTP connections to ESGF, the Copernicus Climate Change Service (C3S), and the Potsdam Institute for Climate Impact Research website.

**Interoperability** is achieved through the use of human- and machine-readable file formats such as CSV, with additional header information (metadata) to ensure coherency and joint data/metadata transfer. These files can be read by most programming languages and software utilities. A few data tables requiring text formatting are distributed in Microsoft's Office Open XML (OOXML) format, which provides open specifications currently implemented in a number of multi-platform office suites. Other data file formats are also open (netCDF, GeoJSON, etc.) or *de facto* standards (e.g. shapefile) provided to facilitate user access. The provided code is also in the multi-platform, free software programming languages R and Python, widely adopted by the community. Code usage examples are provided in Jupyter notebooks, which rely on an open source JSON plain text file format to combine both narrative explanations and code. They can be executed interactively by the open source Jupyter notebook server software in any modern web browser. Their static content is also directly rendered by GitHub. The rest of the files provided in the Atlas repository are also coded in open formats such as PNG (raster images), PDF or SVG (vectorial graphics), or Markdown (formatted plain text files).

**Reusability** is at the core of the Atlas repository design and construct. For this reason, special care has been taken to describe the scope of the provided data, indicating the processing that has been applied to the original data source. This goes to the extent of providing the processing code along with all software versions used, as explained below. All data source versions have been registered, and versioning has also been applied to the data and code produced within the repository. Most of the Atlas repository is licensed under CC-BY 4.0, which promotes maximum reusability allowed by the underlying data sources. Permission has been granted from CMIP6 (the sixth Coupled Model Intercomparison Project, that originally imposed share-alike and some commercial restrictions on their products) to WGI to distribute CMIP6-derived report products, such as the Atlas regionally-aggregated datasets, under a CC-BY-4.0 license.

In addition to the scripts available for reproducibility, several notebooks provide reusable code snippets along with background information, explanations of the code and results, guidance to extend the analyses shown, and references to further information. They are implemented as Jupyter notebooks[8], which are open-source, supporting over 40 different programming languages, and run on a web application which can be accessed locally or from a remote client using any modern web browser. Most of the sample notebooks are provided in R and some in Python.

**Reproducible environment**
Full reproducibility of scientific results requires not only the data and code, but also the whole software environment in which the code was executed. There are several solutions to achieve reproducible execution environments[9]. We adopted incremental Reproducible Execution Environment Specifications (REES) to build a reproducible environment for the Atlas repository. First, we provide the software versions used as a Conda (https://docs.conda.io) environment configuration file and a Docker container configuration file, as well as the resulting Docker image to be readily used. In a local computer, the Conda software and environment manager provides a reliable, multi-platform solution, handling versions and dependencies for most of the software stack. The Docker[10] container configuration file and image extend this by allowing system library dependencies to also be encapsulated. Docker images can be deployed by the user in a local computer, but they also pave the way to deploy the execution environment on cloud computing resources, where virtual machines can be created for particular operating systems running a particular Docker image containing the specific environment used to generate a given scientific result.

In addition, the contents of the Atlas repository can be run interactively on a browser through the MyBinder.org[7] links provided. Binder provides a simple way to compile a Docker image from a wealth of different REES, including Conda configuration files. Moreover, the MyBinder.org initiative provides cloud resources to run interactive notebooks based on Docker images. This provides an execution environment for users to try and modify notebooks, without installing software on their computers. This environment can optionally include additional content from a GitHub repository through the nbgitpuller extension, which in this case is used to retrieve the Atlas repository containing the notebooks and the data they use.

This approach ensures practical reproducibility for most of the contents of the Atlas repository as illustrated in the example described in the Usage Notes section.

## Data records

The Atlas repository is structured in folders including the intermediate products (standard grids, reference regions, etc.) and datasets (see Figure 1). Most of the data inputs come from ESGF[11], which distributes the global and regional climate simulations available from CMIP and the Coordinated Regional Climate Downscaling Experiment (CORDEX), respectively. The first step for findability and accessibility is to build an inventory of all data sources used (described in the *data-sources* folder). This involves parsing metadata for all datasets for scenario and historical experiments in CMIP5, CMIP6 and CORDEX. The final selection of model simulations takes into account the availability of all required variables and analysis periods, discarding erroneous datasets reported in errata registries (e.g. in Earth System Documentation, ES-DOC, errata search service) along with other issues found during the subsequent data processing as documented in the *data-sources* folder of the repository. The final inventory is therefore the result of a semi-automated and iterative process. All issues and decisions made are stored in the *data-sources* folder in the working spreadsheet files used during this process. The final inventory of data sources used, including their version, is also stored as machine-readable CSV files. Moreover, the repository provides unique pointers to the exact data and metadata considered, either as DOIs[12], ESGF search URLs or handle.net[13] URLs. Thus, this folder provides both data provenance information and a machine-readable auxiliary product (the inventory of data sources) to be used in subsequent processing steps.

Another auxiliary product stored in the Atlas repository is a set of reference grids used to bring all data sources to a common spatial resolution. The *reference-grids* folder contains global grid definitions, regular in longitude-latitude, stored in netCDF files containing also common land-sea masks. Commensurable grids (including land-sea masks) are defined starting at 0.25 by 0.25 degrees (used to interpolate ERA5 and observational products) and, by aggregating 2 by 2 grid-cells, grids at 0.5 (CORDEX), 1 (CMIP6), and 2 (CMIP5) degrees resolution were obtained. In addition to these reference grids and masks, there are also other masks used in the Atlas products. These include mountain range masks and missing data masks for several observational products which provide global gridded data but interpolate data from distant stations in empty cells.

A third auxiliary product available at the Atlas repository is the specifications of the new set of AR6 reference regions[14]. This is version 4 of the IPCC WGI reference regions and it builds on the previous version used in AR5[15] with slight modifications to split the most climatically heterogeneous regions, and with the addition of 16 new oceanic regions. The polygons defining the regions are provided in the *reference-regions* folder as CSV files containing corner coordinates and also as shapefile, GeoJSON and R data files. Other sets of regions used in the Atlas are also provided. Namely, IPCC WGII continental regions, 6 land regions subject to monsoons, 28 major river basins, 9 small island regions and another set focused on the analysis of ocean biomes.

The Interactive Atlas Dataset (IAD) consists of a set of 25 climate variables and derived indices used in the Interactive Atlas[16] (see the About section) computed over common grids from several observational datasets plus CMIP5, CMIP6 and CORDEX historical and future projections (under different scenarios) aggregated at a monthly temporal resolution. This data set will be stored by the IPCC Data Distribution Center (https://ipcc-data.org) and is not available as part of the Atlas repository. However, the repository provides the scripts used for the preparation of the IAD (*datasets-interactive-atlas* folder), retrieving the data sources from ESGF, computing derived indices, post-processing (including bias adjustment when necessary) and, finally, interpolating each data set to its reference grid. These scripts rely on the *climate4R* framework[17], version 2.5.3, an open R framework for climate data access and processing. Only interpolation was delegated to the Climate Data Operators (CDO; Schulzweida, 2019), which provides efficient, well-tested, first-order conservative remapping routines, commonly used in climate science.

One of the key resources of the Atlas repository is the direct availability of climate datasets aggregated regionally over the new AR6 WGI reference regions[14] (*datasets-aggregated-regionally* folder). This dataset consists of monthly precipitation and near-surface temperature from the IAD spatially averaged over the reference regions for CMIP5, CMIP6 and CORDEX model output datasets, separately for land, sea, and all (land-sea) grid cells in the region. Regional averages are area-weighted by the cosine of latitude in all cases. This product extends the regional averages presented in Iturbide et al.[14] by including (1) data from CORDEX for all reference regions overlapping the area of the corresponding CORDEX domain by more than 80%, (2) an extended set of CMIP6 models, and (3) for reference, an observation-based product (W5E5[19]) is also provided in the same format, enabling many evaluation studies. These datasets are provided in CSV format as plain data matrices (monthly time series in columns for each AR6 reference region). This format has been extended to include the corresponding metadata as a human- and machine-readable header. Moreover, a sample script (in the R language) used to compute the regional averages and produce these CSV files is also included, along with other functions facilitating their exploitation (e.g. computation of delta changes or depiction of boxplots and scatterplots).

An example of use of the regionally-averaged datasets provided in the repository are the scripts to compute time periods corresponding to different GWLs in the *warming-levels* folder. GWLs are global surface temperature anomalies at a given level (e.g. +1.5ºC) with respect to the average values corresponding to the period 1850-1900 (a proxy for pre-industrial conditions)[20]. GWLs are widely used in policy-making as goals for greenhouse gas (GHG) concentration stabilization. In AR6, the use of GWLs has become common practice to complement scenario-based information on future climate, such as at the end of the century (2081-2100).

Climate information for a given level of global warming can be an effective means of integrating available scenario-based information, independent of when the level of global warming is reached. The warming-level folder within the Atlas repository also provides tables as CSV files and plots with the central year of 20-year periods when different GWLs (+1.5, +2, +3 and +4ºC) are reached for all CMIP5 and CMIP6 Global Climate Model (GCM) simulations considered in the Atlas. Moreover, the scripts and notebooks used to generate these tables and plots are also included. They also allow the creation of new tables and plots for other GWLs or moving-window lengths, using only the repository resources.

Finally, there are two key resources for reproducibility and reusability of the results. The first is the *reproducibility* folder, which contains scripts to reproduce the exact figures shown in the AR6 Atlas chapter. They provide end-to-end sample code to produce the final Atlas figures, providing all processing and plotting steps in all their complexity. The second resource is the *notebooks* folder which provides simpler analyses of the data, illustrating the basic processing workflow on datasets available in the Atlas repository (either the aggregated datasets which are part of the repository or samples of the Interactive Atlas dataset).

## Technical validation
**Atlas FAIR review**
As an integral part of the IPCC AR6, the WGI Atlas chapter was subject to the formal IPCC review process [21]. Additionally, in order to support the integration of FAIR principles in the AR6 and enhance scrutiny of the digital products being developed in the report, the IPCC WGI Technical Support Unit (TSU) launched an informal review process of the data and codes underpinning the Atlas digital products, including the Interactive Atlas. The review was open from 25th November, 2020 until 20th December, 2020 (later extended until 10th January, 2021). This was a pilot to explore, on the one side, how to build on the rigorous review process that IPCC reports undergo and, on another side, how to incorporate open science practices of community scrutiny of digital-based material into the implementation of FAIR principles in the preparation of the Atlas. A selection of 30 reviewers were contacted for this informal review process, gathering expertise from IPCC WGI, WGII and WGIII authors, volunteers from the IPCC TG-Data, and experts from the open data community.

The objective of this review was to gather comments and recommendations about all aspects involved in the implementation of FAIR principles in the Atlas, including the use of platforms and tools (such as GitHub) for open collaboration and version control, the use of open community tools (mainly R) for data processing, the use of metadata standards and model provenance for reproducibility, and the use of Jupyter notebooks to facilitate reusability and provide user guidance on technical aspects. Feedback was also sought on how to improve the repository and the review process, considering how such a review could be extended to all report chapters.

The review was conducted directly on the Atlas repository using the GitHub issue system to collect the reviewers' comments and reply to them. The guidance provided to reviewers suggested addressing some specific questions on some key aspects, amongst others, of the repository (Table 1). A total of 34 issues were opened as a result of this process. The issues are currently closed, but still accessible through GitHub by searching issues labeled "review". This exchange is public and a snapshot of the repository related to the review is accessible through the GitHub tag "v1.6-review".

The main weak points unveiled by this review, that were addressed in the final version of the Atlas, as described in the previous Sections, were:

- Insufficient content description: README files were missing or needed to be extended to provide additional information such as external references for the methodologies applied in the repository, and detailed description of the repository structure in the root README file, which is crucial to guide users.
- Lacking details on the description of the interactive execution environment and execution problems: information on the software versions and the platform requirements were not explicitly included, and the Atlas Hub at some point was not accessible, which was finally deployed through MyBinder.
- Unnecessary content: obscure content of the repository that was not necessary was removed.
- Code cleaning: repetition to be removed, additional explanation comments and headers to be included, and unnecessary and disrupting output elements, such as long warning messages, to be cut down in the notebooks.
- File formats: proprietary formats (e.g. shapefiles) needed to be replaced with alternative open formats.
- Lacking metadata and processing steps: sources of CORDEX and CMIP5 data sets were lacking unique version identifiers, and in the header of CSV files information was missing to fully describe their contents. Some steps in the data processing such as the interpolation software and version were not included.
- Lacking observational data: suggestion was to include also the observational data in the repository, which would be useful for many users to evaluate model results, especially if regridded to the common grid.
- Notebook location and description: it was suggested that notebooks previously scattered across the repository to be gathered in one common location which would be easily accessed by users, and to include sections with more guidance for users to be able to extend their work beyond the examples available in the repository.
- Alternative programming language: As most of the code and notebooks are given in the R programming language, their adaptation to alternative programming languages such as Python would be an added value.
- Licensing and authorship: unclear and mixed licenses needed to be better specified and organized. The suggestion was to release the scripts and notebooks under a Creative Commons license (as the data), instead of the specific General Public License (GPL) software license originally set. Furthermore, the author information was also requested in scripts and notebooks, to be able to comply with the attribution license.

Overall, this informal review greatly improved the Atlas repository in all aspects of the FAIR principles. Open code helps to build community trust, and the review process allowing for comment and potential contribution hopefully translates to a more robust and thorough product than it might otherwise have been.

**Errata and further development**

When the WGI AR6 report was published on August 9$^{th}$ 2021, a final version of the repository was published with the tag "v2.0-final" and a Zenodo snapshot was produced for long-term archival [6]. No further development will take place on this repository since the report content is now frozen with its publication. This means that modifications to the existing resources will take place only for maintenance and to correct for any errata and all changes will be fully documented.

Errata in IPCC reports, such as for chapter figures, need to be documented according to the IPCC Error Protocol. Following the experience of the review described above, error or problem reporting for the Atlas repository related to code and data underpinning figures in the Atlas has been implemented using the GitHub issue tracking system. Moreover, since the Atlas repository supports the implementation of FAIR principles for the Interactive Atlas, the same error or problem reporting system was extended to cover issues reported for this WGI product. This is implemented including a special template with instructions which is linked at the Interactive Atlas landing page. A special tag ("IA") is automatically assigned to these issues. All issues reported are managed by the Atlas support team and classified and labeled as either *errata, problem, suggestion, question* including relevant comments and actions taken (if any). The label *errata* is assigned only to problems affecting the published IPCC reports (e.g. Atlas figures and Interactive Atlas maps and graphics) that also need to be reported to IPCC following the official errata reporting procedure[22].

The Atlas GitHub repository will remain active with basic maintenance and improvement activities, fixing and/or documenting the problems reported (with full tracking on the actions taken) and enhancing reusability with expanded notebooks and examples. These will always be linked to an issue filed in GitHub, explaining the reason to modify the repository and keeping track of these minor changes. The idea is to provide basic maintenance, rather than further develop a repository which is considered final.

Further development of the underlying open tools supporting the Atlas code, in particular the *climate4R* framework for climate data access and post-processing [17], will take place in the original repositories. New data, metadata and sample scripts or notebooks exploiting them can be contributed to this alternative repository, which is also hosted by GitHub.

**Future challenges**

The Atlas repository is a test case developed in the framework of WGI for the comprehensive implementation of FAIR principles into the IPCC report preparation process. This example, together with recommendations issued by the WGI TSU on best practices has raised awareness amongst authors and the IPCC more widely on the value of adopting FAIR principles within the IPCC. Authors of other WGI chapters have also followed the guidance to varying degrees and the outcomes are available on the WGI GitHub repository, with more resources becoming available over time. This implementation experience has helped to refine guidance to support FAIR principles in the IPCC context, develop standard documentation templates and FAIR practices as part of the technical support provided by the TSU to authors.

FAIR principles should be considered as an integral part of future IPCC reports and the efforts required to achieve this should not be underestimated, with adequate human and technical resources allocated to support this task in all chapters, within both the TSU and the chapter author teams. Transparency and reproducibility could be favored by supporting and promoting more widely common infrastructures and resources more widely as part of the preparation of an IPCC report, such as those already deployed by the IPCC-DDC bringing analysis closer to data (Pirani et al. 2022). In addition, leading scientific journals are increasingly requiring authors to make their data and code public, and the IPCC could channel these efforts to be at the forefront of scientific transparency.

A further recommendation is to consolidate the review procedure for code and interactive products that is aligned with the application of FAIR principles and with the IPCC review process. The formal IPCC review process is designed for written reports and is a challenge to implement for code and interactive products. Experts with relevant data and software development knowledge together with expertise in climate science applications need to be involved. Code review could be done in parallel with the formal review of the chapters, but using modern software review tools such as those used in AR6 WGI (GitHub/GitLab), and involving software development communities, such as PANGEO, in the review process. The use of well-established data processing frameworks and standards could help in this process, providing a solid basis for simplifying the code in those tasks for which solutions already exist. In addition, publication of code in open journals that include checklist-driven review (including FAIRness) could also be useful for ancillary packages that implement specific tasks that do not reveal confidential aspects of the assessment; this would be aligned with the publication of scientific results which constitutes the basis of the assessment reports.

## Usage Notes
**Online tools: Example of GWL scaling plots**
In this section, we provide a working example on the use of the Atlas repository resources (code and data) to reproduce and extend some of the results provided by the Interactive Atlas. In particular, we focus on the *Global Warming Level (GWL) scaling plots*, a variant of the empirical scaling relationship, that can be used to quantify the sensitivity of regional changes in different impact-relevant indices as a function of global warming[23]. The analysis in this section has been coded as a new annotated notebook in Python and included in the Atlas repository (GWL-plot_Python.ipynb), expanding reusability and reproducibility. To restrict our analyses to the data available in the Atlas repository, we will focus on regional mean temperature and precipitation. For a given ensemble of climate model projections, the GWL scaling plot represents the decadal mean signals of the regional variable/index of interest versus the corresponding level of global warming, decade by decade (e.g. 2020-2029, 2030-2039, etc.) and ensemble member by member. This visual representation is included in the Interactive Atlas for the analysis of regional information; an example using CMIP6 is provided in Figure 2 displaying regional temperature increases (relative to the 1851-1900 period) in the Mediterranean as a function of the corresponding global warming (for the SSP5-8.5 scenario). In principle, this relationship could be non-linear but, in practice, most variables are shown to linearly scale with global warming, in most of the regions and robustly across scenarios [23]. Ensemble averages are also shown in this plot for each decade as overlaid dark red dots. These allow us to easily read the evolving mean ensemble global warming and regional response from the corresponding axes.

For this exercise, we recreated the Interactive Atlas GWL plot using Python code (Figure 3) to exploit the resources provided by the Atlas repository. Moreover, we extend the plot provided by the Interactive Atlas by adding the linear scaling factor (β) and the coefficient of determination ($R^2$), as a measure of the linearity of the relationship. These are derived from a standard least squares fit, also shown in the new figures. Finally, the line representing a unit scaling factor is added for reference. Figure 3 compares mean temperature GWL plots for CMIP6 and CMIP5 for 3 different regions. The companion notebook available at the notebooks folder of the repository provides the code and guidance to reproduce these figures and to reuse the code for other variables (e.g. precipitation), regions and model ensembles (e.g. CORDEX).

Overall, Figure 3 shows similar scaling properties (similar slope β) between CMIP5 and CMIP6. The main difference being the known [24] tendency of CMIP6 towards higher climate sensitivity. For the last decade of the century, CMIP6 models reach global surface air temperatures beyond 7ºC, while CMIP5 remains below 6ºC. Also, for some regions such as northeastern Northamerica (NEN) there is a difference in multi-model spread between the two CMIP ensembles, affecting the goodness of fit for the linear relationship ($R^2$). Finally, northern Europe (NEU), seems to be prone to outlier models in the regional response, which could be due to different sea ice extent. The Interactive Atlas can be used to easily identify those outliers --EC-Earth3-Veg and NESM3 (CMIP6) and IPSL-CM5B-LR (CMIP5)--, since this tool shows on hover the model corresponding to each point in the scatter.

## Code Availability

All code, notebooks and datasets in the Atlas repository are available at https://github.com/IPCC-WG1/Atlas under the Creative Commons Attribution license, CC-BY 4.0 (with the exception of the aggregated data corresponding to a few CORDEX simulations which are distributed under CC-BY-NC 4.0). The notebook illustrating GWL plots is available at https://github.com/IPCC-WG1/Atlas/blob/devel/notebooks/GWL-plot_Python.ipynb

## Acknowledgements

We acknowledge partial funding from projects ATLAS (PID2019-111481RB-I00) funded by MCIN/AEI/10.13039/501100011033 and IS-ENES3 which is funded by the European Union's H2020 programme under grant agreement No 824084. We also acknowledge the World Climate Research Programme's Working Group on Coupled Modelling and Working Group on Regional Climate, responsible for CMIP and CORDEX, respectively. We also thank the climate modeling groups for producing and making available their model output, as described in the data-source folder of the repository. We also acknowledge the Earth System Grid Federation infrastructure, an international effort led by the U.S. Department of Energy's Program for Climate Model Diagnosis and Intercomparison, the European Network for Earth System Modelling and other partners in the Global Organisation for Earth System Science Portals (GO-ESSP). The opinions expressed are those of the author(s) only and should not be considered as representative of the European Commission's official position.

## Author contributions
Anna Pirani, Özge Yelekçi, José M. Gutiérrez, Alaa Al Khourdajie and David Huard contributed

to the framing of the FAIR principles in the context of the Technical Support Units and Task Group on Data Support for Climate Change Assessments (TG-Data) and the definition of the review protocol. Maialen Iturbide, Jesús Fernández, José M. Gutiérrez, Joaquin Bedia, Ezequiel Cimadevilla, Javier Diez-Sierra, Rodrigo Manzanas, Ana Casanueva, Jorge Baño-Medina, Josipa Milovac, Sixto Herrera, Antonio S. Cofiño, Daniel San-Martín and Markel García-Díez have contributed to the creation of the contents of the Atlas repository. Alaa Al Khourdajie, Matteo De Felice, Javier Diez-Sierra, Jesus Fernandez, James Goldie, Dimitris A. Herrera, Rodrigo Manzanas, Aparna Radhakrishnan, Alessandro Spinuso, Kristen Thyng and Claire Trenham reviewed and shaped the final version of the Atlas repository. Jesus Fernandez created the figures. Maialen Iturbide, Jesus Fernandez and José M. Gutiérrez wrote the first draft. All authors contributed to the final writing and revision of the manuscript.

## Competing interests

The authors declare no competing interests.

# Figures

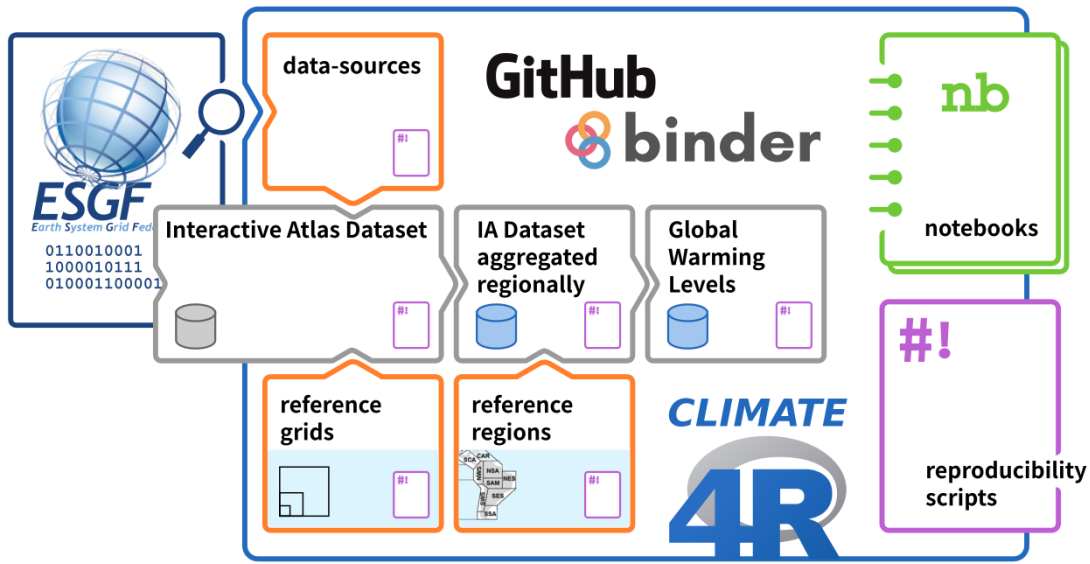

**Figure 1:** Schematic representation of the contents of the GitHub Atlas repository backing the Interactive Atlas and Atlas chapter figures. Data at different stages of processing are shown in grey boxes, representing different directories in the GitHub repository. Blue (grey) cylinders represent datasets (not) readily available in the repository. Orange boxes are auxiliary information used at different stages. ESGF provides metadata to build the data sources catalogue, which dictates the data to be processed, also retrieved from ESGF. Notebooks and scripts, mainly relying on the *climate4R* framework, feed from intermediate datasets at different stages to illustrate their use and reproduce specific Atlas chapter figures, respectively. All directories contain a subdirectory with scripts allowing to reproduce the corresponding contents. To enable a reproducible execution environment, the repository contents are loaded in a MyBinder cloud environment with all the required software pre-installed.

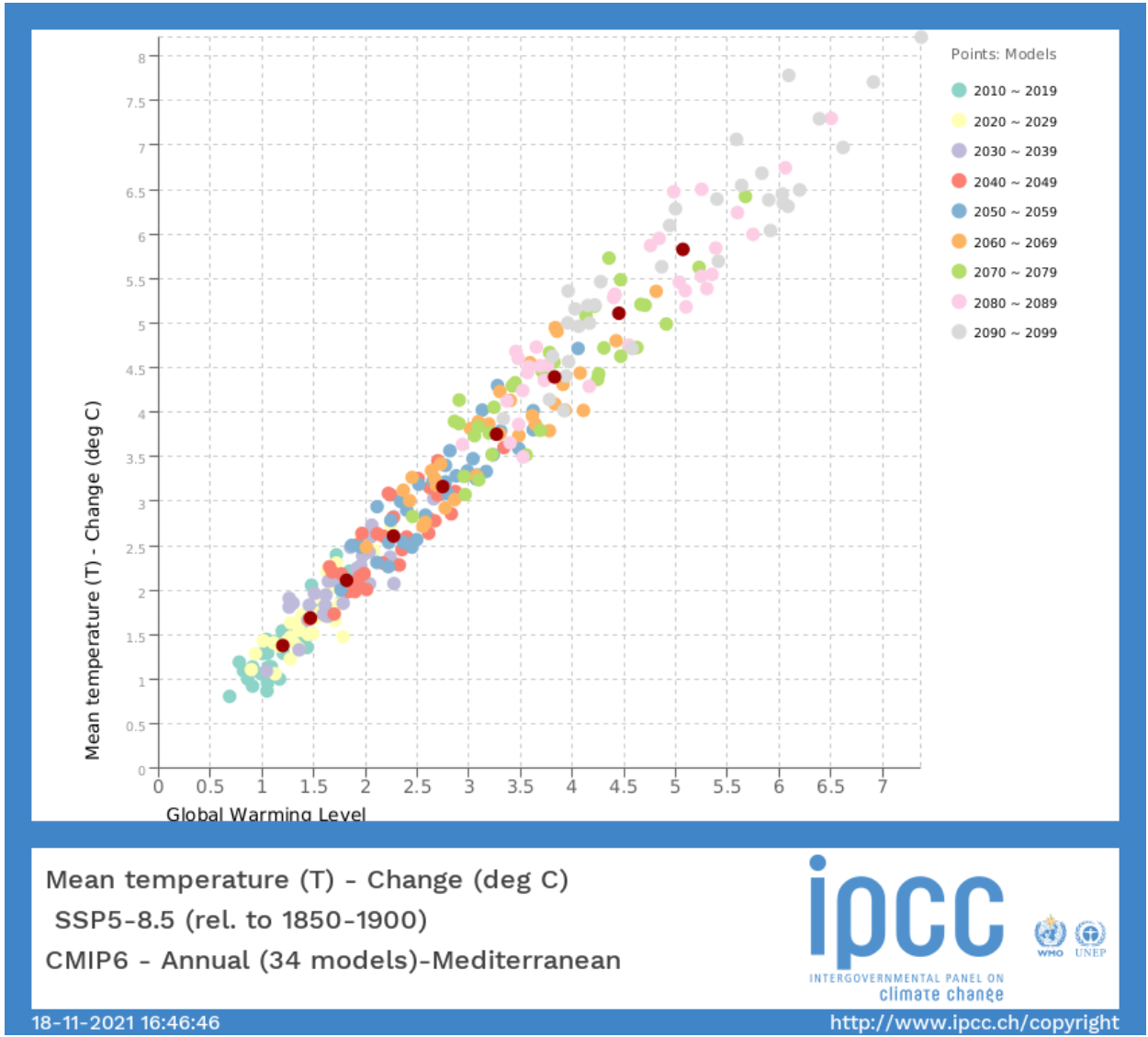

**Figure 2:** Near surface temperature CMIP6 GWL scaling plot for the Mediterranean region (MED). Source: IPCC WGI AR6 Interactive Atlas [16].

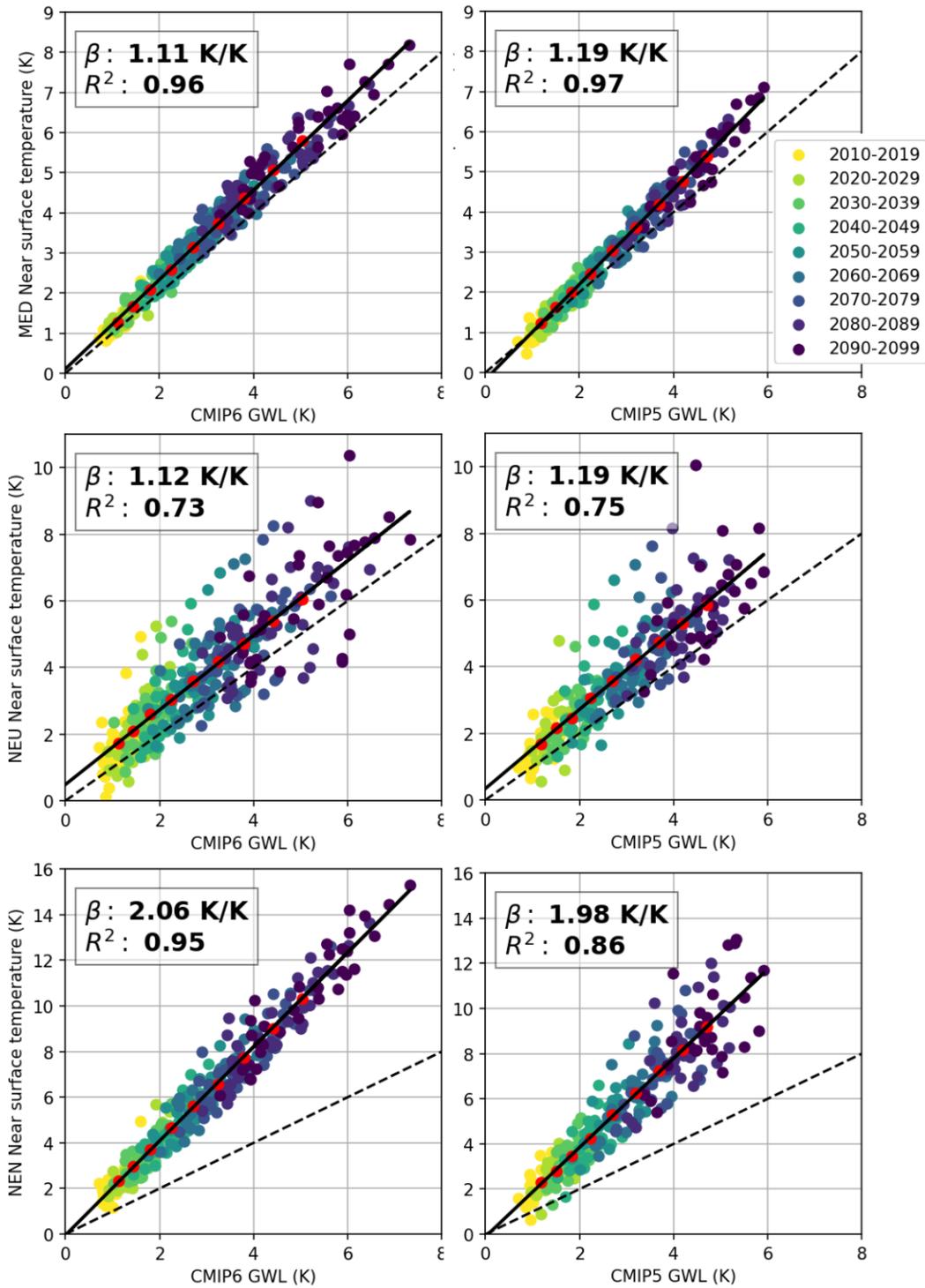

**Figure 3:** Near surface temperature GWL scaling plots for (top) the Mediterranean, (middle) Northern Europe, and (bottom) Northeastern Northamerica regions for CMIP6 (left) and CMIP5 (right) ensemble projections (SSP5-8.5 and RCP8.5 scenarios, respectively).

# Tables

**Table 1:** Questions and key aspects to be addressed that were suggested in the revision process of the Atlas repository.

| Technology | Do you find the used community tools appropriate/useful (in particular GitHub and R)? |
|---|---|
| Structure | Do you find the structure of the repository appropriate in terms of navigation and functionality? |
| Scope | Is the scope of the repository (datasets, reference regions and reference periods – including warming levels) clearly described? |
| Guidance | Does the Atlas repository describe clearly the structure and use? What information is missing and what improvements could be made? |
| Provenance | Do you find the provenance and intermediate information provided (including, e.g. model metadata and warming levels) suitable? |
| Usefulness | Do you find this repository useful as a resource to the scientific community? |
| Reusability | Do you find the R scripts and notebooks well documented and easy to follow (in particular for code reusability)? |
| Notebooks | Do you find the auxiliary Jupyter notebooks informative and useful? |
| License | Do you find the license for the code appropriate? |
| Development | Do you have ideas for further development of the repository? |